\documentclass[journal]{IEEEtran}
\usepackage{multirow}

\usepackage{xcolor}
\usepackage{ifpdf}
\usepackage{amsfonts}
\usepackage{cite}
\ifCLASSINFOpdf
 \usepackage[pdftex]{graphicx}
\else
  \usepackage[dvips]{graphicx}
\fi
\usepackage{amsmath}
\usepackage{array}
\DeclareMathOperator{\Tr}{Tr}

\begin{document}

\title{Localization based on enhanced low frequency interaural level difference}

\author{Metin Calis,
Steven van de Par,
Richard Heusdens and
Richard C. Hendriks
\thanks{This project was supported in part by the Deutche Forschungsgesellschaft (DFG), Grant No. EXS 2177: Hearing4all and the National Science Agenda (NWA) idea generator project: Restored sound localization for hearing impaired people, NWA.1228.191.034.
}
\thanks{ M. Calis, R. C. Hendriks, and R. Heusdens are with the Faculty of Electrical Engineering, Mathematics and Computer Science, Delft University of Technology 2628 CD Delft, The Netherlands (e-mail: m.calis@tudelft.nl; r.c.hendriks@tudelft.nl; r.heusdens@tudelft.nl).}
\thanks{S. Par is with the Department of Medical Physics and Acoustics, Carl von Ossietzky University Acoustics Group 26111 Oldenburg, Germany  (e-mail: steven.van.de.par@uol.de).
}}

\maketitle

\begin{abstract}
The processing of low-frequency interaural time differences is found to be problematic among hearing-impaired people. The current generation of beamformers does not consider this deficiency. In an attempt to tackle this issue, we propose to replace the inaudible interaural time differences in the low-frequency region with the interaural level differences. In addition, a beamformer is introduced and analyzed, which enhances the low-frequency interaural level differences of the sound sources using a near-field transformation. The proposed beamforming problem is relaxed to a convex problem using semi-definite relaxation. The instrumental analysis suggests that the low-frequency interaural level differences are enhanced without hindering the provided intelligibility. A psychoacoustic localization test is done using a listening experiment, which suggests that the replacement of time differences into level differences improves the localization performance of normal-hearing listeners for an anechoic scene but not for a reverberant scene. 
\end{abstract}


% \fi
\IEEEpeerreviewmaketitle

\section{Introduction}
Good hearing is a vital and important part of daily life. Being hearing impaired comes with many challenging situations ranging from private to social interactions. In some cases, hearing-impaired people can find themselves in dangerous situations due to the lack of hearing. For example, when crossing in traffic. Moreover, hearing-impaired people might feel isolated in practical situations due to the inability to differentiate and understand different sound sources in complex listening environments \cite{bronkhorst2015cocktail}. %Moreover, the correlation between aging and hearing loss proves the prevalence of the research on improving the quality of hearing \cite{Glyde2013TheEO}. 
%Around $11$\% of the Dutch population suffers from hearing loss and around $22$\% of Europe is estimated to have some degree of hearing loss \cite{shieldb.2006}.
These people can benefit by using hearing aids. However, despite being very powerful, the current generation of hearing aids is not able to completely compensate for hearing loss. 

The current generation of hearing aids comes with a wireless link that enables the left and right microphone arrays to exchange information to achieve noise reduction and \textcolor{black}{conservation of spatial auditory cues} \cite{kates2008digital}. The microphone arrays are then combined and used in beamforming algorithms to enhance the target source while suppressing the interferers. Many beamforming algorithms have been proposed, such as the binaural minimum variance distortionless response beamformer \cite{sesliandreas} and the joint binaural linearly constrained minimum variance beamformer \cite{hadadtheoretical}. The former is known to be good at achieving maximum noise reduction, while the latter is known for its ability to \textcolor{black}{preserve the spatial cues} and \textcolor{black}{reduce the noise} at the same time. Besides quality and intelligibility, in many practical situations, it is also important for the suppressed interferers to sound natural and to appear from their original locations. For example, to correctly localize (interfering) point sources in the surrounding as in a traffic scenario. This leads to the typical trade-off between noise reduction and preservation of the spatial cues as examined by researchers, which led to many beamforming algorithms \cite{koutrouvelis2016relaxed} \cite{doclo2006theoretical} \cite{ARCHERBOYD201864}.

%To understand how existing beamforming algorithms tend to preserve the localization information and investigate how this can be further improved, some basic understanding is required on human sound localization. 
The sound source signals reaching the two ears contain the information that is required for the auditory system to extract and analyze the horizontal location of the sources. The auditory events are formed through two main dissimilarities that exist between the signals reaching the ears; namely, interaural time differences and interaural level differences \cite{blauert1997spatial}. The interaural time differences are the primary source of localization for frequencies below $1.5$ kHz and mainly occur due to the differences in the time that takes for a source to reach both ears. The interaural level difference on the other hand is caused by the shadowing of the head and exploited mainly for frequencies above $3$ kHz \cite{Hartmann1999HowWL}. The frequencies between these ranges are the frequencies where none of the spatial cues \textcolor{black}{are} dominant and considered to be the range that is the worst for localization \cite{Hartmann1999HowWL}. The aforementioned beamformers that preserve the location of the interferers, mainly preserve the time and level differences of the interferer after processing. In some cases, the spatial cues that are preserved are in fact not audible by hearing-impaired people \cite{bardsley2019use} \cite{Lorenzi18866} \cite{evalthebenefitofhaoncock}. This paper is part of a project that aims at exploring the suprathreshold effects of why the hearing impaired can not achieve similar performance to normal hearing people even after the audibility is established. In an attempt to answer this question, it has been hypothesized that hearing-impaired people can achieve a better performance when the auditory scene is presented in a different acoustic form than the original. 

The low-frequency processing of ITDs, which is known in the literature as the temporal fine structure processing (TFS) is found to be problematic among hearing-impaired people \cite{grose2010processing}. Several reasons have been proposed in the literature to understand why low-frequency TFS processing can be impaired. The phase-locking of the auditory-nerve fibers might have become imprecise \cite{miller1997effects} or the phase shift on the basilar membrane might result in inaccurate decoding of the TFS information. The TFS information is thought to be dependent on the cross-correlation across the different places on the basilar membrane \cite{ruggero1994cochlear}. If there is a phase shift, the cross-correlation would give inaccurate TFS information reducing its reliability. Another reason proposed by the authors in \cite{broadened} suggests that the broadening of the auditory filters might be the cause of the reduced TFS processing. In the light of these suggestions, the beamformers that preserve the binaural cues of the whole spectrum might be wasting the degrees of freedom that they have, to preserve the inaudible cues. Instead of designing beamformers that disregard the hearing impairment of the user, a better enhancement algorithm can be created by taking into account how well the listener utilizes the preserved binaural cues.

This paper starts with the signal model and the background information in Section \ref{sec:signal} and Section \ref{sec:background}, respectively. After that, the related work is explained in Section  \ref{sec:related}. The problem formulation is explained in Section \ref{sec:problem}. The following chapters cover the solution to the problem that is explained earlier. These chapters include the applied convex relaxation and parameter selection. In Section \ref{sec:performance}, the proposed method is analyzed using theoretical measures. This section includes objective intelligibility tests and localization analysis. An experiment using normal listeners has been conducted and the results are shown in Section \ref{sec:experiment}. The paper 
ends with a discussion and a conclusion chapter where the findings are analyzed and comments have been made about the future work that can be done.
\section{Signal Model}
\label{sec:signal}
In our signal model, we assume the presence of two hearing aids with $M/2$ microphones at each ear, where $M$ is an even number. The binaural enhancement algorithms apply spatial filters to the left and the right hearing aid microphones, where each filter uses the recordings of all $M$ microphones together. It is assumed that there is one target signal and there are $r$ interferers where the maximum number of interferers is represented as $r_{\rm{max}}$. The enhancement is applied in the frequency domain where $k$ represents the frequency bin and $l$ represents the frame index. The signal received by the $j$th microphone for $j=1,\ldots,M$ is given by
\begin{align}
\label{eq:signalmodel}
    y_j(k,l)= \underbrace{a_j(k,l)s(k,l)}_{x_j(k,l)}+\sum_{i=1}^r\underbrace{b_{ij}(k,l)u_i(k,l)}_{n_{ij}(k,l)}+v_j(k,l)
\end{align}
where
\begin{itemize}
    \item $s(k,l)$ denotes the target signal at the location of the source,
    \item $u_i(k,l)$ denotes the $i$th interferer signal at the location of the source,
    \item $a_j(k,l)$ is the acoustic transfer function (ATF) of the target signal with respect to the $j$th microphone,
    \item $b_{ij}(k,l)$ is the ATF of the $i$th interfering signal from the location of the source to the $j$th microphone,
    \item $v_j(k,l)$ is the additive noise at the $j$th microphone.
\end{itemize}

In practice, the acoustic transfer functions (ATF) are estimated  \cite{markovich2015performance}. However, in this work we assume the ATF to be known to avoid errors due to the ATF mismatch. In the instrumental performance evaluation in Section \ref{sec:performance}, the ATFs are generated according to a predefined setup and in the perceptual evaluation in Section \ref{sec:experiment} generic head-related transfer functions (HRTF) are used. 

To simplify the notation, the indices $k$ and $l$ are omitted. The signal model in (\ref{eq:signalmodel}) can then be written in vector notation as,

\begin{align}
    \mathbf{y}=\underbrace{\mathbf{a}s}_{\mathbf{x}}+\sum_{i=1}^r\underbrace{\mathbf{b}_iu_i}_{\mathbf{n}_i}+\mathbf{v}
    \end{align}
where $\mathbf{a}\in \mathbb{C}^{M \times 1}$, $\mathbf{b}_{ij}\in \mathbb{C}^{M \times 1}$, $\mathbf{y}\in \mathbb{C}^{M \times 1}$, $\mathbf{x}\in \mathbb{C}^{M \times 1}$, $\mathbf{n}_i \in \mathbb{C}^{M \times 1}$ and $\mathbf{v}\in \mathbb{C}^{M \times 1}$ are the column vectors that hold the $M$ realizations of $a_j$, $b_{ij}$, $y_j$, $x_j$, $n_{ij}$ and $v_j$ for $j=1\ldots M$ and $i=1\ldots r$ respectively.

The noise and the sound sources are assumed to be mutually uncorrelated. Using this assumption, the cross power spectral density (CPSD) of the received signal at the $j$th microphone can be written as,

\begin{align}
    \mathbf{P}_\mathbf{y}=E[\mathbf{y}\mathbf{y}^H] = \mathbf{P}_\mathbf{x}+\underbrace{\sum_{i=1}^r\mathbf{P}_{\mathbf{n}_i}+\mathbf{P}_\mathbf{v}}_{\mathbf{P_N}}
\end{align}
where
\begin{itemize}
    \item $\mathbf{P}_\mathbf{x}=E[\mathbf{x}\mathbf{x}^H]\in \mathbb{C}^{M \times M}$ is the CPSD matrix of $\mathbf{x}$.
    \item $\mathbf{P}_{\mathbf{n}_i}=E[\mathbf{n}_i\mathbf{n}_i^H]\in \mathbb{C}^{M \times M}$ is the CPSD matrix of $\mathbf{n}_i$.
    \item $\mathbf{P}_\mathbf{v}=E[\mathbf{v}\mathbf{v}^H] \in \mathbb{C}^{M \times M}$ is the CPSD matrix of $\mathbf{v}$,
    \item $\mathbf{P_{N}}=\sum_{i=1}^r\mathbf{P}_{\mathbf{n}_i}+\mathbf{P}_\mathbf{v}\in \mathbb{C}^{M \times M}$ is the CPSD matrix of the total noise.
\end{itemize}

All the applied beamformers in this paper use the ideal voice activity detection and overcome any estimation errors which might happen in practice. 

Without loss of generality, the last microphone and the first microphone are selected as the left reference microphone and the right reference microphone for the target. These are denoted as $a_L$ and $a_R$, respectively. For the interferers, the left and right microphones are denoted relative to the $i$th interferer as $b_{i,L}$ and $b_{i,R}$ for $i=1,\ldots,r$.

The incidence angle is assumed to be $0$ degrees at the mid-sagittal plane and increases clockwise to $180^\circ$ degrees and decreases to $-180^\circ$ anticlockwise starting from the median plane.
\section{Background Information}
\label{sec:background}
In this section, background information is given that will be used in the following chapters. The information consists of the binaural cues and the binaural beamforming algorithms.
\subsection{Binaural Cues}
The auditory system uses the time and level differences of the signals reaching the ears to determine the horizontal location of the sound \cite{blauert1997spatial}. Following the convention in \cite{Koutrouvelis_2017} and \cite{DocloCoherence}, the time and level differences can be defined using the interaural transfer function (ITF). The input and output ITF for the target signal is defined as 
\begin{align}
    ITF_x^{in}=\frac{a_L}{a_R} \,,\quad ITF_x^{out}=\frac{\mathbf{w}_L^H \mathbf{a}}{\mathbf{w}_R^H \mathbf{a}}\,.
\end{align}
The ITF for the interferer is defined as
\begin{align}
    ITF_{u_i}^{in}=\frac{b_{i,L}}{b_{i,R}} \,,\quad ITF_{u_i}^{out}=\frac{\mathbf{w}_L^H \mathbf{b}}{\mathbf{w}_R^H \mathbf{b}}\,.
\end{align}
The interaural level difference is defined to be the squared magnitude of the ITF. The ILD before processing can be represented as
\begin{align}
\label{eq:ILD}
ILD_x^{in}=|ITF_x^{in}|^2\,,\quad ILD_{u_i}^{in}=|ITF_{u_i}^{in}|^2 \,.
\end{align}
The output ILD is defined similarly, that is,
\begin{align}
ILD_x^{out}=|ITF_x^{out}|^2\,,\quad ILD_{u_i}^{out}=|ITF_{u_i}^{out}|^2 \,.
\end{align}
The interaural time differences are defined to be the phase of the ITF normalized by the angular frequency, $w$ \cite{ITDayari}. For the input ITD, this can be shown as
\begin{align}
\label{eq:ITD}
    ITD_x^{in}=\frac{\angle ITF_x^{in}}{2 \pi f}\,,\quad  ITD_{u_i}^{in}=\frac{\angle ITF_{u_i}^{in}}{2 \pi f} \,.
\end{align}
Similarly, the output ITD is defined as
\begin{align}
    ITD_x^{out}=\frac{\angle ITF_x^{out}}{2\pi f}\,,\quad  ITD_{u_i}^{out}=\frac{\angle ITF_{u_i}^{out}}{2\pi f} \,.
\end{align}

In some cases, the interaural time differences are represented using the phase information \cite{AndreasAnalysis}. The interaural phase differences for the sources before processing can then
be defined as
\begin{align}
    \label{eq:IPD}
    IPD_x^{in}=\angle ITF_x^{in}\,,\quad  IPD_{u_i}^{in}=\angle ITF_{u_i}^{in} \,.
\end{align}
After the processing, the IPDs can be written as
\begin{align}
    IPD_x^{out}=\angle ITF_x^{out} \,,\quad  IPD_{u_i}^{out}=\angle ITF_{u_i}^{out} \,.
\end{align}
If the beamformer preserves the ITF of the source after processing, then both the ILD and ITD will also be preserved. On the other hand, a beamformer might be written such that at specific frequencies only the ILD or ITD cues are preserved, discarding the other binaural cue.  
\subsection{Binaural Beamforming Algorithms}
Some beamformer algorithms such as the binaural minimum variance distortionless response (BMVDR) aim at reducing the noise power as much as possible while other beamformers such as the joint binaural linearly constrained minimum variance algorithm (JBLCMV) aim at achieving noise reduction while preserving the spatial cues of the interferers. 
The BMVDR improves the listening comfort by providing the maximum noise suppression without preserving any of the spatial cues of the interferers. All the binaural cues of the interferers collapse to the target's direction forcing the interferers to sound from the same direction as the target. The optimization problem of the BMVDR is represented by the following expression, that is,
\begin{align}
    \min_{\mathbf{w}_L,\mathbf{w}_R \in \mathbb{C}^{Mx1}}& \mathbf{w}_L^H \mathbf{P_N} \mathbf{w}_L + \mathbf{w}_R^H \mathbf{P_N} \mathbf{w}_R\nonumber\\
    \text{s.t.} \quad &\mathbf{w}_L^H \mathbf{a} = a_L\,, \quad \mathbf{w}_R^H \mathbf{a}=a_{R} \,.
    \label{eq:bmvdr}
\end{align}

The optimization problem given in (\ref{eq:bmvdr}) has the following closed form solutions \cite{docloassistivedevices} \cite{hadadtheoretical}, which can be expressed as
\begin{align}
    \hat{\mathbf{w}}_L = \frac{\mathbf{P_N}^{-1}\mathbf{a}a_L^*}{\mathbf{a}^H\mathbf{P_N}^{-1}\mathbf{a}}\,, \quad \hat{\mathbf{w}}_R = \frac{\mathbf{P_N}^{-1}\mathbf{a}a_R^*}{\mathbf{a}^H\mathbf{P_N}^{-1}\mathbf{a}}\,.
    \label{eq:bmvdrw}
\end{align}
If the left and right spatial filters $\mathbf{w}_L$ and $\mathbf{w}_R$ are merged into one vector $\mathbf{w}_{\rm{BMVDR}}=[\mathbf{w}_L^H \quad \mathbf{w}_R^H]^H$, the optimization problem (\ref{eq:bmvdr}) can be written jointly as
\begin{align}
    \min_{\mathbf{w}\in \mathbb{C}^{2Mx1}} & \mathbf{w}^H \Tilde{\mathbf{P}}\mathbf{w}\\
    \text{s.t.} \quad &\mathbf{w}^H\boldsymbol{\Lambda}_A=\mathbf{f}_A^H\,,
\end{align}
where
\begin{align}
    \mathbf{\Tilde{P}}&=\begin{bmatrix} \mathbf{P_N} &\mathbf{0}\\ \mathbf{0} & \mathbf{P_N}\end{bmatrix}\,,\quad
    \boldsymbol{\Lambda}_A&= \begin{bmatrix}\mathbf{a} & \mathbf{0} \\ \mathbf{0} & \mathbf{a}\end{bmatrix}\,,\quad
    \mathbf{f}_A &= \begin{bmatrix}a_L \\ a_R \end{bmatrix}\,.
\end{align}
The closed-form solution to the jointly written BMVDR problem  \cite{hadadtheoretical} is given by
\begin{align}
\label{eq:bmvdrsol}
    \mathbf{w}_{\rm{BMVDR}}=\Tilde{\mathbf{P}}^{-1}\mathbf{\boldsymbol{\Lambda}_A}(\mathbf{\boldsymbol{\Lambda}_A}^H\Tilde{\mathbf{P}} \mathbf{\boldsymbol{\Lambda}_A})^{-1}\mathbf{f}_A\,.
\end{align}
When (\ref{eq:bmvdrsol}) is used, all the interfering sound sources will sound from the same location as the target source. This prevents a possible increase in the intelligibility due to the spatial release from masking (SRM) \cite{kidd2010stimulus}. The SRM suggests that the sounds that are spatially separated have higher intelligibility then sources that are colocated. The JBLCMV beamformers solve this problem by introducing additional constraints to preserve the spatial information of the interferers. This comes up with a trade-off with regards to possible noise reduction which can be provided since the feasible set of the spatial filters to perform the noise reduction will shrink with the additional constraints. 

The JBLCMV framework is given by the following formulation,
\begin{align}
\label{eq:jblcmv}
   \min_{\mathbf{w}\in \mathbb{C}^{2Mx1}} & \mathbf{w}^H \Tilde{\mathbf{P}}\mathbf{w}\\
    \text{s.t.} \quad &\mathbf{w}^H\boldsymbol{\Lambda}=\mathbf{f}^H\,,
\end{align}
where
\begin{align}
    \boldsymbol{\Lambda}&=\left[\begin{array}{c|c}
        \boldsymbol{\Lambda}_a & \boldsymbol{\Lambda}_b  
    \end{array}\right] \\
    &=\left[\begin{array}{c c| c c c}
    \mathbf{a} & \mathbf{0} & \mathbf{b}_1b_{1R} & \dots & \mathbf{b}_mb_{mR}\\
    \mathbf{0} & \mathbf{a} & -\mathbf{b}_1b_{1L} & \dots & -\mathbf{b}_mb_{mL}\end{array}\right] \in \mathbb{C}^{2M \times (2+r)}\,, \\
    \mathbf{f}^H &= \left[\begin{array}{c|c}
        \mathbf{f}^H_a & \mathbf{f}^H_b  
    \end{array}\right] \\
    &= \left[\begin{array}{c c| c c c c} 
    a^*_L & a^*_R & 0 &0&\dots & 0 
    \end{array}\right]\in \mathbb{C}^{1\times(2+r)}\,.
\end{align}
In addition to the target distortionless constraints $\mathbf{w}^H\boldsymbol{\Lambda}_a=\mathbf{f}_a$, there are additional constraints $\mathbf{w}^H\boldsymbol{\Lambda}_b=\mathbf{f}_b$, which preserve the cues of the interferers. These additional constraints preserve the ITF of the interferers forcing both ITD and ILD to be preserved in the whole spectrum.

Assuming that there are $r$ interferers and $M$ microphones, the degrees of freedom that are left for the JBLCMV to do noise reduction is $2M-2-r$. Here the distortionless target constraint reduces the total degrees of freedom by two whereas, interferer cue preservation reduces the total degrees of freedom by $r$ as there is one equality constraint per interferer. On the other hand, the BMVDR has $2M-2$ degrees of freedom left to do noise reduction as there are only two target distortionless constraints. This enables the BMVDR to have a larger domain to minimize noise power which provides a better noise reduction capability. For relaxed methods such as presented in \cite{sesliandreas}, the degrees of freedom are not straightforward due to the inequality constraints. However, by checking the feasible set of the optimization problem, a comparison can still be done. The authors of \cite{sesliandreas} relax the interferer cue preservation constraint providing a user-controlled trade-off between noise reduction and cue preservation. The relaxation provides a feasible set between the JBLCMV and the BMVDR. Hence, the output noise power becomes bounded between the JBLCMV and the BMVDR.
\section{Related Work}
\label{sec:related}
%How we hear, 
In \cite{moore2016evaluation}, a method to enhance the low-frequency ILDs was introduced. The authors first solved the phase ambiguity problem, which might occur for frequencies below $1500$ Hz. The resulting unambiguous ITD values were plugged into the ILD-to-ITD function measured for high-frequency tones which were obtained from \cite{feddersen1957localization}. The resulting ILDs were smoothed and \textcolor{black}{sent} to the bilateral hearing aids to replace the low-frequency ITD cues. The authors did not find any improvement in intelligibility. However, they found an improvement of localization for the speech stimulus but not for the broadband noise, lowpass filtered noise, or lowpass filtered AM noise. 

In \cite{francart2009amplification} and \cite{francart2011enhancement}, the authors analyzed the localization performance of bimodal listeners when the ILD in the available dynamic range \textcolor{black}{was} enhanced. In the former, the signals \textcolor{black}{were} noise-vocoded at one ear and lowpass filtered at the contralateral ear to simulate the bimodal hearing. The authors first measured the ILD of the full-band received signals on the hearing aid and cochlear implant devices and used the root-mean-square ratio of the signals to enhance the low-frequency content. It \textcolor{black}{was} found that the localization ability of the normal-hearing listeners under a simulated hearing setup \textcolor{black}{was} improved for a broadband noise by $14^\circ$ but not for the telephone alerting signal. In the latter study, the authors created an artificial ILD versus angle function which \textcolor{black}{overcame} the non-monotonicity of the ILD signals around $60^\circ$. This \textcolor{black}{was} done by using white noise as the source and the resulting natural ILD versus angle functions of the six bimodal listeners' hearing devices that were placed on a mannequin's head. This function \textcolor{black}{was} obtained by using the full-band spectrum of the received signal which \textcolor{black}{was} used irrespective of the stimulus spectrum. The resulting ILD-to-angle function \textcolor{black}{was} transformed into a monotonic relation by extrapolating the curve at the point when the non-monotonicity \textcolor{black}{started}. The authors report\textcolor{black}{ed} a $4^\circ$ to $10^\circ$ improvement for the horizontal localization performance for the bimodal listeners. Following this line of work, the authors in \cite{dieudonne2018head} designed a beamformer that attenuated the sources coming from the contralateral direction as opposed to traditional beamformers which attenuated the sources coming from the rear. The authors reported an improvement of horizontal localization and speech intelligibility for bimodal listeners. 

Spatial release from masking (SRM) using the enhanced low-frequency ILD cues \textcolor{black}{was} investigated by \cite{rana}. Spatial release from masking is the improvement in speech reception thresholds when the target and the distractor change from the same to different locations. Normal hearing people benefit around 20 dB SRM, whereas the hearing impaired benefit less \cite{rana}. Apart from the horizontal separation which benefits the intelligibility of the signals, in \cite{shinn}\cite{jin2011spatial} the authors realized that distance cues also improve the SRM. In the former, one sound source was fixed to one meter and the other source \textcolor{black}{was} moved closer to the listener. It \textcolor{black}{was} found that a better target-to-masker ratio \textcolor{black}{could} be achieved for the better ear. In the latter, three experiments \textcolor{black}{were} conducted which \textcolor{black}{assessed} the effect of distance cues on spatial segregation. \textcolor{black}{It was} found out that the intelligibility of the target source \textcolor{black}{could} be improved due to the spatial release from masking. Following this line of work, the authors in \cite{rana} investigated the use of low-frequency ILDs separately on the spatial release from masking. Maximum low-frequency ILDs \textcolor{black}{were} applied and it was found out that HI can benefit from an additional increase of SRM.

This work investigates the effect of near-field low-frequency distance cues on horizontal localization. It is different from \cite{moore2016evaluation} because of the formulation of how the enhanced ILD cues are generated. In addition to this, a low-frequency enhancing beamformer is proposed and its performance is analyzed. The non-monotonicity of the ILD cues \textcolor{black}{is} avoided by limiting the low-frequency range that is going to be used for ILD enhancement, unlike \cite{francart2011enhancement} where an artificial angle versus ILD function is generated empirically. 
\section{Problem Formulation}
\label{sec:problem}
The duplex theory proposed by \cite{rayleigh} suggests that the ITD cues are effective for frequencies below $1500$ Hz and ILD cues are effective for frequencies above $1500$ Hz. Although these ranges are perceptually the most important frequencies for the processing of pure tone ILD and ITD cues, the sensitivity of the human ear to these binaural cues extends beyond this scope. In the high frequencies, human listeners have been observed to be sensitive to the slowly-varying envelope fluctuations of broadband signals \cite{bernstein1994detection}. In the low frequencies, the sensitivity to the fast-varying temporal fine structure of the signal is observed \cite{moore2008role}. The sensitivity to the ILD is observed in the whole region.

\begin{figure}
    \centering
    \includegraphics[width=3.5in]{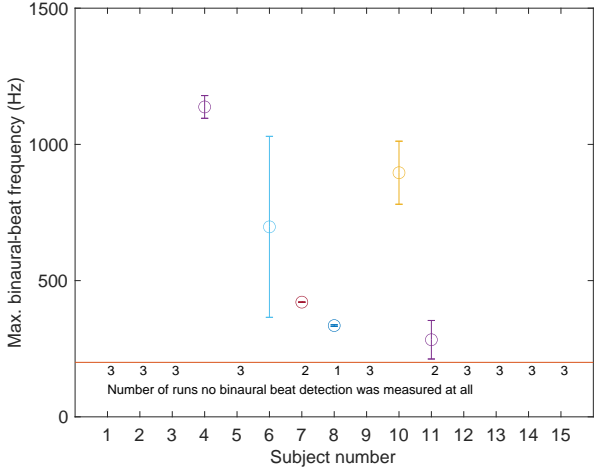}
    \caption{The mean and the standard deviation of the maximum binaural beat frequency that the panel of hearing-impaired people can hear.}
    \label{fig:binauralbeat}
\end{figure}

The deficiency of low-frequency ITD processing up to $1000$ Hz \textcolor{black}{was} observed among hearing-impaired people \cite{TFS1} \cite{TFS2} \cite{TFS3}, where the authors found a correlation between increasing age and the inability to process the low-frequency temporal fine structure. \textcolor{black}{During the preparation of this paper, a binaural beat listening experiment with a panel of 15 hearing-impaired people was carried out. The stimuli used in the experiment consisted of two tones with a slightly different frequency in the left and right ear, approximately around 2 Hz. The subjects were presented with three intervals, one of which contained the binaural beat stimulus with different frequencies in the left and right ear, while the other intervals contained identical sinusoids with some randomized frequency to make sure that the differences in pitch could not be used. Using an adaptive staircase method, the frequency of the tones was increased until the beat stimulus could not be heard. It was found out that 9 out of 15 subjects were unable to hear a binaural beat even at a low frequency like 200 Hz, while all but one subjects could not hear binaural beats in the frequency range that normal hearing people can still hear binaural beats. The results can be seen in Fig. \ref{fig:binauralbeat}}.

In the frequency range where the temporal fine structure cues are dominant, it is still possible to differentiate the ILD cues. The just noticeable difference of ILD is approximately $1$ dB in the whole spectrum \cite{blauert1997spatial}. The ILD in the low-frequency region can still be heard and claimed to be the dominant cue for distance perception \cite{brungart1998near}. These ILD cues do not occur naturally for the far-field signals. In the far-field, the ILD cues are around $0$ dB for frequencies below $500$ Hz \cite{bosunxie}. Since the diameter of the head is significantly smaller than the wavelength, the head does not shadow the contralateral ear and hence, level differences do not occur. 

In the near-field region defined to be less than $1$ m \cite{brungart1998near}, low-frequency ILDs can reach as high as 20-30 dB. When a source comes closer to the head, the level differences in the whole spectrum increase, including the low-frequency region. As head shadowing is not observed in this region, the near-field ILDs and the source angle create an injective relation which can be utilized to boost the horizontal localization performance.

Binaural beamformer algorithms such as the binaural linearly constrained minimum variance algorithms spend the degrees of freedom that they have to preserve both the ILD and ITD cues in the whole spectrum \cite{van1988beamforming}. If the ITD cues are not heard by the hearing-impaired person, the degrees of freedom which can be used for noise reduction will be wasted for preserving the inaudible ITD cues. Since the low-frequency ITD cues are not processed by the hearing impaired, we will investigate whether they can artificially be replaced by the near-field ILD cues in the same region. This research project, therefore, aims to answer the following  question: \textit{Can the enhancement of low-frequency ILDs overcome the loss of spatial information induced by the deficient temporal fine structure processing?}

\section{Proposed Method}
%Explain the JBLCMV at high freq, and ILD at low freq.
In this section, a beamformer is introduced to enhance the low-frequency ILDs of the interferers while keeping the target distortionless. The section starts with a brief overview of the problem, continues with the applied convex relaxation and the methodology behind choosing a particular scaling factor that determines to which extend the low-frequency ILDs are enhanced. 

In the low-frequency region, the interaural level differences will be artificially introduced, while in the high frequency region, both the ILD and ITD will be preserved. This can be expressed with the following optimization problem:
\begin{equation}
\label{eq:mainProblem}
\begin{aligned}
f<f_{c}
    \min_{\mathbf{w}_L,\mathbf{w}_R \in \mathbb{C}^{Mx1}}& \mathbf{w}_L^H\mathbf{P}\mathbf{w}_L+\mathbf{w}_R^H\mathbf{P}\mathbf{w}_R\\
    \text{s.t.} \quad &\mathbf{w}_L^H \mathbf{a} = a_L \quad \mathbf{w}_R^H \mathbf{a}=a_{R}\,, \\
    &\left|\frac{\mathbf{w}_L^H \mathbf{b_i}}{\mathbf{w}_R \mathbf{b_i}}\right|^2-c\left|\frac{b_{i,L}}{b_{i,R}}\right|^2=0\,,
\\ \\
f>f_c
        \min_{\mathbf{w}_L,\mathbf{w}_R \in \mathbb{C}^{Mx1}}& \mathbf{w}_L^H\mathbf{P}\mathbf{w}_L+\mathbf{w}_R^H\mathbf{P}\mathbf{w}_R\\
    \text{s.t.} \quad &\mathbf{w}_L^H \mathbf{a} = a_L \quad \mathbf{w}_R^H \mathbf{a}=a_{R}\,,\\
    &\mathbf{w}_L^H\mathbf{b}_ib_{iR}-\mathbf{w}_R^H\mathbf{b}_ib_{iL}=0\,,
\end{aligned}
\end{equation}
for  $i=1,...,r'\leq r_{\rm{max}}$.
This means that for all the frequencies above the cut-off frequency $f_c$, the JBLCMV will be applied which is explained in (\ref{eq:jblcmv}). For the frequencies below $f_c$, the proposed method will be applied which enhances the low-frequency ILD while leaving the target undistorted. The enhancement of the ILDs is represented by the last constraint
\begin{align*}
\left|\frac{\mathbf{w}_L^H \mathbf{b_i}}{\mathbf{w}_R^H \mathbf{b_i}}\right|^2-c\left|\frac{b_{i,L}}{b_{i,R}}\right|^2=0. 
\end{align*}
Without the scaling factor $c$, the input ILD of each interferer will be preserved at the output. By including the factor $c$, we aim at enhancing the ILDs of the interferers. This factor will depend on the direction of the interferer and the near-field transformation. The sound sources that are close to the head have a higher ILD and the sound sources that are away from the head have a lower ILD in the low-frequency region. In addition to this, the magnitude of the ILDs increases as the sound source reaches to $90$ or $-90$ degrees starting from the mid-sagittal plane. A description of this relation for different frequencies can be seen in Fig. \ref{fig:quberlin}.

While there is a closed-form solution for the JBLCMV, there is no closed-form solution for the introduced low-frequency enhancement beamformer due to its non-convexity. In the following section, the optimization problem for $f<f_c$ will be relaxed using a semi-definite relaxation approach \cite{vandenberghe1996semidefinite}.
\subsection{Convex Relaxation}
\label{subsec:convex}

The problem at hand preserves the target signal at the left and right ears. In addition to that, the ILD of the interferers has been enhanced by the scaling factor $c$.

The ILD enhancement constraint in (\ref{eq:mainProblem}) can be extended and written as 
\begin{equation}
\label{eq:expansionInterferer}
\begin{aligned}
    0&=\left|\frac{\mathbf{w}_L^H \mathbf{b_i}}{\mathbf{w}_R^H \mathbf{b_i}}\right|^2-c\left|\frac{b_{i,L}}{b_{i,R}}\right|^2 \, \\
    &=\mathbf{w}_L^H\mathbf{b_i}\mathbf{b_i}^H\mathbf{w}_L|b_{i,R}|^2 - c\mathbf{w}_R^H\mathbf{b}_i\mathbf{b_i}^H\mathbf{w}_R|b_{i,L}|^2\, \\
    &=\underbrace{\begin{bmatrix}
        \mathbf{w}_L^H & \mathbf{w}_R^H  
    \end{bmatrix}}_{\mathbf{w}^H \in \mathbb{C}^{2M \times 1}}
    \underbrace{\begin{bmatrix}
        \mathbf{b}_i\mathbf{b}_i^H|b_{i,R}|^2 & \mathbf{0} \\
        \mathbf{0} & -c\mathbf{b}_i\mathbf{b}_i^H|b_{i,L}|^2 
    \end{bmatrix}}_{\mathbf{M}_i\in \mathbb{C}^{2M \times 2M}}
    \underbrace{\begin{bmatrix}
        \mathbf{w}_L \\ \mathbf{w}_R   
    \end{bmatrix}}_{\mathbf{w}}\,.
\end{aligned}
\end{equation}
Using the expansion in (\ref{eq:expansionInterferer}), the main problem given in (\ref{eq:mainProblem}) can be written compactly as
\begin{equation}
\label{eq:mainProblem2}
\begin{aligned}
    \min_{\mathbf{w}\in \mathbb{C}^{2M \times 1}} &  \underbrace{\begin{bmatrix}
        \mathbf{w}_L^H & \mathbf{w}_R^H  
    \end{bmatrix}}_{\mathbf{w}^H \in \mathbb{C}^{1 \times 2M}}
    \underbrace{\begin{bmatrix}
    \mathbf{P} & \mathbf{0} \\ 
    \mathbf{0} & \mathbf{P}
    \end{bmatrix}}_{\mathbf{\Tilde{P}} \in \mathbb{C}^{2M \times 2M}}
     \underbrace{\begin{bmatrix}
        \mathbf{w}_L \\ \mathbf{w}_R  
    \end{bmatrix}}_{\mathbf{w}\in \mathbb{C}^{2M \times 1}} \\
     \text{s.t.}\quad & \underbrace{\begin{bmatrix}
        \mathbf{w}_L^H & \mathbf{w}_R^H  
    \end{bmatrix}}_{\mathbf{w}^H }\underbrace{\begin{bmatrix} \mathbf{a} & \mathbf{0} \\ \mathbf{0} & \mathbf{a}\end{bmatrix}}_{\boldsymbol{\Lambda}_a \in \mathbb{C}^{2M \times 2}} = \underbrace{\begin{bmatrix}
    a_L & a_R \end{bmatrix}}_{\mathbf{f}_a^H \in \mathbb{C}^{1 \times 2}} \,, \\
    &\underbrace{\begin{bmatrix}
        \mathbf{w}_L^H & \mathbf{w}_R^H  
    \end{bmatrix}}_{\mathbf{w}^H}
    \mathbf{M}_i
\underbrace{\begin{bmatrix}
    \mathbf{w}_L \\ \mathbf{w}_R   
\end{bmatrix}}_{\mathbf{w}}=0 \,.
\end{aligned}
\end{equation}
If vector notation is used, the optimization problem given in (\ref{eq:mainProblem2}) can be written as
\begin{equation}
\label{eq:mainproblemVectorized}
    \begin{aligned}
        \min_{\mathbf{w}} &\quad\mathbf{w^H}\mathbf{\Tilde{P}}\mathbf{w}\\
        \text{s.t.}&\quad  \mathbf{w}^H\boldsymbol{\Lambda}_a=\mathbf{f}_a^H\,,\\
        &\quad\mathbf{w}^H\mathbf{M}_i\mathbf{w}=0\,.
    \end{aligned}
\end{equation}
The optimization problem in (\ref{eq:mainproblemVectorized}) is not convex due to the quadratic equality constraint $\mathbf{w}^H\mathbf{M}_i\mathbf{w}$. Since $\mathbf{M_i}$ is not positive semi-definite, the expression $\mathbf{w}^H\mathbf{M}_i\mathbf{w}$ is not convex. On the other hand, the objective function is convex as $\mathbf{\Tilde{P}}$ is positive semi-definite. The linear equality $\mathbf{w}^H\boldsymbol{\Lambda}_a=\mathbf{f}_a^H$ is convex. If the equality constraint at the last line of (\ref{eq:mainproblemVectorized}) is written as two inequality constraints,
the problem at hand becomes a Quadratically Constrained Quadratic Program (QCQP), which is NP-hard \cite{park2017general}. We can use semi-definite relaxation \cite{SDRrelaxationOfQCQP}  and reformulation-linearization techniques \cite{qualizza2012linear} to transform the optimization problem from (\ref{eq:mainproblemVectorized}) to a relaxed convex problem which can be solved in polynomial time using interior-point solvers such as CVX in MATLAB. 

The semi-definite relaxation can be applied to transform (\ref{eq:mainproblemVectorized}) into a semi-definite program (SDP). Two important matrix properties will be used to execute the transformation. 
 \begin{enumerate}
  \item We have the following relation for any quadratic expression
  \begin{align}
      \mathbf{q}^H\mathbf{Z}\mathbf{q}=\Tr(\mathbf{q}^H\mathbf{Z}\mathbf{q})=\Tr(\mathbf{q}\mathbf{q}^H\mathbf{Z})\,.
    \label{eq:quadratic}
  \end{align}
  \item Using Schur's complement \cite{GoluVanl96},
  \begin{equation}
      \begin{aligned}
      \mathbf{Z}=\begin{bmatrix}
      \mathbf{A} & \mathbf{B} \\
      \mathbf{B}^H & \mathbf{C}
      \end{bmatrix}\geq0 \Leftrightarrow\\
      \mathbf{A}\geq 0\,,\quad (\mathbf{I}-\mathbf{A}\mathbf{A}^\dag)\mathbf{B}=\mathbf{0}\,,\quad \mathbf{S}_1\geq 0\,,\\
        \mathbf{C}\geq 0\,,\quad (\mathbf{I}-\mathbf{C}\mathbf{C}^\dag)\mathbf{B}^H=\mathbf{0}\,,\quad \mathbf{S}_2\geq 0\,,\\
      \end{aligned}
      \label{eq:shur}
  \end{equation}
  
\end{enumerate}
with $\mathbf{S}_1=\mathbf{C}-\mathbf{B}^H\mathbf{A}^\dag\mathbf{B}$ the generalized Schur complement of $\mathbf{A}$ in $\mathbf{Z}$ and $\mathbf{S}_2=\mathbf{A}-\mathbf{B}\mathbf{C}^\dag\mathbf{B}^H$ the generalized Schur complement of $\mathbf{C}$ in $\mathbf{Z}$. $\mathbf{A}^\dag$ is the pseudo-inverse of $\mathbf{A}$.

Let $\mathbf{W}=\mathbf{w}\mathbf{w}^H$. Using (\ref{eq:quadratic}), the optimization problem in (\ref{eq:mainproblemVectorized}) becomes
\begin{equation}
    \begin{aligned}
    \text{Problem\;1:}\quad\quad
    \min_{\mathbf{w},\mathbf{W}}&\Tr(\mathbf{W}\Tilde{\mathbf{P}})\\
    \text{s.t.}&\quad  \mathbf{w}^H\boldsymbol{\Lambda}_a=\mathbf{f}_a^H \,,\\
    &\Tr(\mathbf{W}\mathbf{M}_i)=0\,,\quad i=1,\ldots,r\,,\\
    &\mathbf{W}=\mathbf{w}\mathbf{w}^H\,.
\end{aligned}
    \label{eq:OPTstep1}
\end{equation}

The optimization problem given in (\ref{eq:OPTstep1}) is not convex due to the last equality constraint $\mathbf{W}=\mathbf{w}\mathbf{w}^H$. Removing this rank 1 constraint makes the problem convex \cite{vandenberghe1996semidefinite},
\begin{equation}
    \begin{aligned}
    \min_{\mathbf{w},\mathbf{W}}&\Tr(\mathbf{W}\Tilde{\mathbf{P}})\\
    \text{s.t.}&\quad  \mathbf{w}^H\boldsymbol\Lambda_a=\mathbf{f}_a^H \,,\\
    &\Tr(\mathbf{W}\mathbf{M}_i)=0\,,\quad i=1,\ldots,r\,,\\
    &\mathbf{W}\geq\mathbf{w}\mathbf{w}^H\,.
    \label{eq:OPTstep2}
\end{aligned}
\end{equation}
Finally using (\ref{eq:shur}), (\ref{eq:OPTstep2}) can be written in the standard SDP form as,
\begin{equation}
\begin{aligned}
    \text{Problem\;2:}\quad\quad
    \min_{\mathbf{w},\mathbf{W}}&\Tr(\mathbf{W}\Tilde{\mathbf{P}})\\
    \text{s.t.}&\quad  \mathbf{w}^H\boldsymbol\Lambda_a=\mathbf{f}_a^H \,,\\
    &\Tr(\mathbf{W}\mathbf{M}_i)=0\,,\quad i=1,\ldots,r\,,\\
    &\begin{bmatrix}
    \mathbf{W} & \mathbf{w} \\
    \mathbf{w}^H & \mathbf{1}
    \end{bmatrix}\geq 0\,.
    \end{aligned}
    \label{eq:OPTstep3}
\end{equation}

The optimization problem in (\ref{eq:OPTstep3}) is convex and can be solved in polynomial time. Due to the relaxation introduced to the equality constraint $\mathbf{W}=\mathbf{w}\mathbf{w}^H$, the new problem in (\ref{eq:OPTstep3}) does not necessarily give the same solution as (\ref{eq:OPTstep1}). Let the optimal arguments be $\Hat{w}^*$ and $\Hat{W}^*$. If $\Hat{W}^*=\Hat{w}^*\Hat{w}^{*^H}$, the solution to (\ref{eq:OPTstep3}) is the optimal value for (\ref{eq:OPTstep1}), which is equal to the solution of the original problem in (\ref{eq:mainProblem}).

Let the optimal value for (\ref{eq:OPTstep1}) be $p_1^*$ and the optimal value for (\ref{eq:OPTstep3}) be $p_2^*$. Due to the relaxation that is introduced, the feasible set of (\ref{eq:OPTstep3}) is larger than the feasible set of (\ref{eq:OPTstep1}). For this reason, $p_2^*$ lower bounds $p_1^*$ as there is a larger set for the minimization problem in (\ref{eq:OPTstep3}) which encapsulates the set of (\ref{eq:OPTstep1}). Using reformulation-linearization techniques, we can further tighten this bound by introducing redundant constraints to the problem in (\ref{eq:OPTstep2}). The target distortionless equality constraints in (\ref{eq:OPTstep2}) can be reformulated using \cite{qualizza2012linear} as follows. The target distortionless constraint is given by
\begin{align}
    \mathbf{w}^H\boldsymbol\Lambda_a=\mathbf{f}_a^H\,.
    \label{eq:targetdistortionless}
\end{align}
If we multiply left and right by $\mathbf{w}$, (\ref{eq:targetdistortionless}) becomes
\begin{align}
    \mathbf{w}\mathbf{f}_a^H &= \mathbf{w}\mathbf{w}^H\boldsymbol\Lambda_a\nonumber\\
    \mathbf{w}\mathbf{f}_a^H &= \mathbf{W}\mathbf{\Lambda}_a\,. \label{eq:targetdistortionlessrlt}
\end{align}
In addition to (\ref{eq:targetdistortionlessrlt}), another redundant constraint can be added to tighten the bound even further. The target distortionless constraint in (\ref{eq:targetdistortionless}) can be reformulated as,
\begin{align}
    0 &= \mathbf{w}^H\boldsymbol\Lambda_a-\mathbf{f}_a^H\nonumber\\
     &= (\mathbf{w}^H\boldsymbol\Lambda_a-\mathbf{f}_a^H)(\mathbf{w}^H\boldsymbol\Lambda_a-\mathbf{f}_a^H)^H \nonumber \\
     &= \mathbf{w}^H(\mathbf{\Lambda_a}\boldsymbol\Lambda_a^H)\mathbf{w}-\mathbf{w}^H\mathbf{\Lambda}_a\mathbf{f}_a-\mathbf{f}_a^H\mathbf{\Lambda}_a^H\mathbf{w}+\mathbf{f}_a^H\mathbf{f}_a \nonumber \\
     &= \Tr(\mathbf{W}\mathbf{\Lambda}_a\mathbf{\Lambda}_a^H)-\mathbf{w}^H\mathbf{\Lambda}_a\mathbf{f}_a-\mathbf{f}_a^H\mathbf{\Lambda}_a^H\mathbf{w}+\mathbf{f}_a^H\mathbf{f}_a \,. \label{eq:targetdistortionlessrlt2}
\end{align}
If the feasible set of (\ref{eq:OPTstep2}) is tightened by the addition of (\ref{eq:targetdistortionlessrlt}) and (\ref{eq:targetdistortionlessrlt2}), the optimization problem becomes
\begin{equation}
    \begin{aligned}
    \text{Problem\;3:}\quad
        \min_{\mathbf{w},\mathbf{W}}&\Tr(\mathbf{W}\Tilde{\mathbf{P}})\\
    \text{s.t.}&\quad  \mathbf{w}^H\boldsymbol\Lambda_a=\mathbf{f}_a^H \,,\\
    &\mathrm{Tr}(\mathbf{W}\mathbf{M}_i)=0\,,\\
    &\begin{bmatrix}
    \mathbf{W} & \mathbf{w} \\
    \mathbf{w}^H & \mathbf{1}
    \end{bmatrix}\geq 0\,,\\
    &\Tr(\mathbf{W}\mathbf{\Lambda}_a\mathbf{\Lambda}_a^H)-\mathbf{w}^H\mathbf{\Lambda}_a\mathbf{f}_a-\\&\mathbf{f}_a^H\mathbf{\Lambda}_a^H\mathbf{w}+\mathbf{f}_a^H\mathbf{f}_a=0\,,\\
    &\mathbf{W}\mathbf{\Lambda}_a=\mathbf{w}\mathbf{f}_a^H\,,
\end{aligned}
\label{eq:OPTstep4}
\end{equation}
for $i=1\ldots r\leq r_{\rm{max}}$.
Let the optimal value for the optimization problem in (\ref{eq:OPTstep4}) be $p^*_3$. Due to the additional constraints, we have
\begin{align}
\label{eq:overestimate}
    p_1^*\geq p_3^*\geq p_2^* \,.
\end{align}
The feasible set is tightened and the approximation is improved. 
\subsection{Cut-off Frequency}
\label{subsec:cutoff}

\textcolor{black}{The dynamic range of the ILD cues in the near field is significantly larger than the ILD cues in the far field for frequencies below $1000$ Hz. For example, at $500$ Hz, the ILD increases from 4 dB to 19 dB as a source at $90^\circ$ approaches the head from $1$ m to $0.12$ m} \cite{brungart1998near}. We can introduce these low-frequency ILD cues artificially in an attempt to overcome the lack of low-frequency ITD processing, which can be seen among hearing-impaired people \cite{king2014effects} \cite{moore2014auditory}.

The injective nature of the ITDs provides a reliable cue for frequencies below $1500$ Hz. This reliability has to be provided if the ITDs are intended to be replaced by the ILDs.

\begin{figure}[!t]
    \centering
    \includegraphics[width=3.5in]{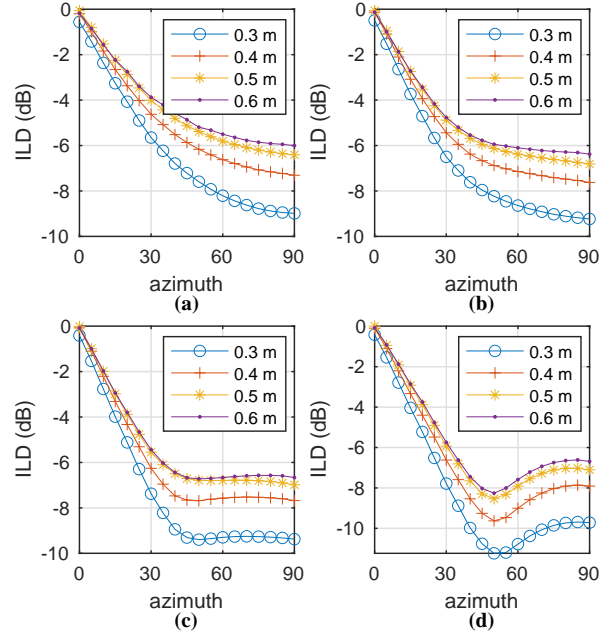}
    \caption{ILD versus azimuth plotted using the HRTF taken from \cite{quberlin} for four different frequencies (a) ILDs at 800 Hz (b) ILDs at 1000 Hz (c) ILDs at 1200 Hz, (d) ILDs at 1500 Hz}
    \label{fig:quberlin}
\end{figure}

To understand which frequency should be used as a cut-off frequency for the beamformer algorithms, an analysis is done using the head-related related transfer functions recorded using a KEMAR manikin at the Technical University of Berlin \cite{quberlin}. Four distances are selected: $0.3$, $0.4$, $0.5$, and $0.6$ meters. The ILDs are calculated using four frequencies: $800$, $1000$, $1200$ and $1500$ Hz. The results can be seen in Fig. \ref{fig:quberlin}. It is observed that for frequencies $1200$ and $1500$ Hz, the head shadow is visible and the ILD to frequency function is not injective. For $1000$ Hz, the ILD is monotonic but a flattening in the same range can be observed. For this reason, $800$ Hz is selected to be the cut-off frequency for the low-frequency ILD enhancement. 
%\begin{figure}[!t]
%\centering
%\includegraphics[width=2.5in]{myfigure}
% where an .eps filename suffix will be assumed under latex, 
% and a .pdf suffix will be assumed for pdflatex; or what has been declared
% via \DeclareGraphicsExtensions.
%\caption{Simulation results for the network.}
%\label{fig_sim}
%\end{figure}

\subsection{Scaling Factor}
\label{subsec:scalingfactor}
The scaling factor $c$ in (\ref{eq:mainProblem}) can be decided through a look-up table which can be generated using the near-field head-related transfer functions \cite{quberlin} \cite{bosunxie} \cite{kolnhrtf}. Instead, the spherical model introduced by \cite{rabinowitz1993sound} and later verified by \cite{duda} can be used to generate the near-field distance cues from the far-field HRTF. This method has been proposed by \cite{kan2009} and has been shown psychoacoustically to give accurate estimates. 

\begin{figure}[!t]
    \centering
    \includegraphics[width=3.5in]{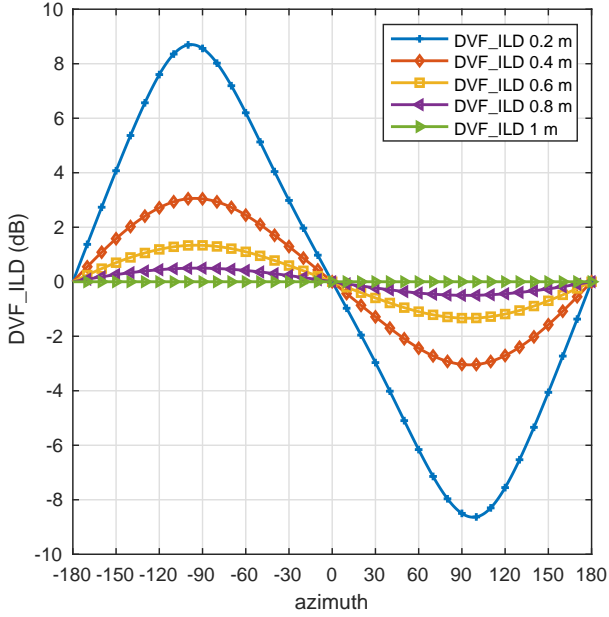}
    \caption{The scaling factor $DVF\_ILD$ calculated for 5 different distances.}
    \label{fig:DVFILD}
\end{figure}

The distance variation function (DVF) proposed by \cite{kan2009} assumes the head can be considered as a rigid sphere with a radius $a$. The ears are assumed to be located at $100^\circ$ away from the mid-sagittal plane. The rigid sphere model estimates the pressure at a point on the surface of a sphere as
\begin{align}
    p(a,w,\theta,r)=-kr\sum_{m=0}^\infty(2m+1)\frac{h_m(kr)}{h_m'(ka)}P_m(cos(\theta))e^{-ikr}\,,
\end{align}
where $h_m(kr)$ is a spherical Hankel function of the first kind of order $m$ and $h'_m(ka)$ is the first derivative at radius $a$, $k=\frac{w}{c}$ is the wavenumber, $c$ is the speed of sound, $P_m(\Lambda)$ is the Legendre polynomial of degree $m$, $\theta$ is the angle between a vector that starts from the center of the sphere and ends at the source location and the vector that starts at the center of the sphere and ends at a location on the surface of the sphere, $r$ is the distance of the source to the center of the sphere and $w$ is the angular frequency \cite{rabinowitz1993sound}. The DVF is calculated as 
\begin{align}
    DVF=\frac{p_n(\alpha,w,\theta,d_n)}{p_f(\alpha,w,\theta,d_f)}\,,
\end{align}
where $p_n$ stands for the near field pressure on the surface of the sphere and $p_f$ stands for the pressure on the surface of the sphere for a far-field source. Although the authors used an individualized head radius, an average radius $8.75$ cm is used in this study. If the $DVF$ is calculated at the desired frequencies, the near-field HRTF can be calculated by 
\begin{align}
    HRTF(d_n)=DVF \times HRTF(d_f)\,,
    \label{eq:DVF}
\end{align}
where $d_n$ is the near-field distance and $d_f$ is the far-field distance.
In addition to this, the far-field ILD cues can be transformed into near-field ILD cues by using (\ref{eq:DVF}). Given that the left ear head-related transfer function is named as $HRTF_L(d)$ and the right ear head-related transfer function as $HRTF_R(d)$, the ILDs can be calculated as
\begin{align}
    HRTF_L(d_n) &= DVF_L \times HRTF_L(d_f)\,,\nonumber\\
    HRTF_R(d_n) &= DVF_R \times HRTF_R(d_f)\,,\nonumber\\
    \underbrace{\frac{HRTF_L(d_n)^2}{HRTF_R(d_n)^2}}_{ILD_n} &= \underbrace{\frac{DVF_L^2}{DVF_R^2}}_{DVF\_ILD}\times\underbrace{\frac{HRTF_L(d_f)^2}{HRTF_R(d_r)^2}}_{ILD_f}\nonumber\,,\\
    ILD_n &= DVF\_{ILD} \times ILD_f\,,
    \label{eq:DVFILD}
\end{align}
where $ILD_n$ stands for the near-field ILDs, $ILD_f$ stands for the far-field ILDs, and $DVF\_ILD$ stands for the scaling factor, which relates the near-field ILD cue to the far-field ILD cue. The scaling factor $DVF\_ILD$ is calculated until $800$ Hz for the distances $0.2$ m, $0.4$ m, $0.6$ m, $0.8$ m, $1$ m and shown in Fig. \ref{fig:DVFILD}. Since the far-field distance is assumed to be $1$ m, $DVF\_ILD$ is $0$ dB for 1 m. It reaches in magnitude to $8$ dB for a source at $0.2$ m and at $90$ degrees to the left or right of the mid-sagittal plane. 
The final problem becomes 
\begin{equation}
\label{eq:mainProblem3}
\begin{aligned}
\text{f$<$800:} \quad \quad \quad \;\;
    \min_{\mathbf{w},\mathbf{W}}&\Tr(\mathbf{W}\Tilde{\mathbf{P}})\\
    \text{s.t.}&\quad  \mathbf{w}^H\boldsymbol\Lambda_a=\mathbf{f}_a^H \,,\\
    &\Tr(\mathbf{W}\mathbf{M}_i)=0\,,\\
    &\begin{bmatrix}
    \mathbf{W} & \mathbf{w} \\
    \mathbf{w}^H & \mathbf{1}
    \end{bmatrix}\geq 0\,,\\
    &\Tr(\mathbf{W}\mathbf{\Lambda}_a\mathbf{\Lambda}_a^H)-\mathbf{w}^H\mathbf{\Lambda}_a\mathbf{f}_a-\\&\mathbf{f}_a^H\mathbf{\Lambda}_a^H\mathbf{w}+\mathbf{f}_a^H\mathbf{f}_a=0\,,\\
    &\mathbf{W}\mathbf{\Lambda}_a=\mathbf{w}\mathbf{f}_a^H\,, 
\\ \\
\text{f$\geq$800:}
        \min_{\mathbf{w}_L,\mathbf{w}_R \in \mathbb{C}^{Mx1}}& \mathbf{w}_L^H\mathbf{P}\mathbf{w}_L+\mathbf{w}_R^H\mathbf{P}\mathbf{w}_R\\
    \text{s.t.} \quad &\mathbf{w}_L^H \mathbf{a} = a_L \quad \mathbf{w}_R^H \mathbf{a}=a_{R}\,,\\
    &\mathbf{w}_L^H\mathbf{b}_ib_{iR}-\mathbf{w}_R^H\mathbf{b}_ib_{iL}=0\,,
\end{aligned}
\end{equation}
where $c=DVF\_ILD$, which is generated using (\ref{eq:DVFILD}), $f$ stand for the frequency and $i=1\ldots r\leq r_{\rm{max}}$. In the following sections, the optimization problem given in (\ref{eq:mainProblem3}) is represented with the abbreviation $ILD_{d}$ where $d$ stands for the near field distance. For example, if enhancement with respect to $0.2$ m is used, it is represented as $ILD_{0.2}$.

\section{Performance Measures}
\label{sec:performance}
In this section, the intelligibility and the localization performance of the proposed algorithm \textcolor{black}{are} investigated using objective measures. For the intelligibility, the speech intelligibility in bits metric is used \cite{van2017instrumental}. For the spatial cue preservation performance, the absolute value of the difference between the input cue and the output cue is used.

We create a synthetic scenario with 8 interferers, where the first five are selected to be random sentences from the TIMIT database speech corpus \cite{timit} and the others are different noise types such as non-stationary noise, speech shaped noise, and babble noise. The interferer $u_1$, $u_2$, $u_3$, $u_4$, $u_5$, $u_6$, $u_7$ and $u_8$ are located at $-20^\circ$, $20^\circ$, $ -40^\circ$, $40^\circ$, $-60^\circ$, $60^\circ$, $-90^\circ$, $90^\circ$ degrees respectively. There is one target signal $s$, which is located at $0^\circ$. 

The anechoic head-related function from the Oldenburg database \cite{kayser2009database} is used to generate the acoustic transfer functions. The sampling frequency is selected to be $16$ kHz to cover \textcolor{black}{the most energetic components for speech signals}. Each signal is windowed with a square-root-Hann window with a frame length of $12.5$ ms and $50$\% overlap. The signals are concatenated such that there is a $20$ second of voice active signal duration. For the application of the beamformer algorithms, a 256-point short-term fast Fourier transform (STFT) is applied to each frame. After processing the signals in the frequency domain, each frame is converted back to the original time domain by multiplying with a square-root Hann window and taking the inverse Fourier transform. The frames are overlapped accordingly and the time domain signal is obtained. 

Each interferer is scaled to be $0$ dB with respect to the target signal so that cues related to audibility can be avoided. In addition, white gaussian noise at an SNR of $50$ dB with respect to the target source is added to the received signals after they have been processed by the respective acoustic transfer functions to imitate the microphone's self-noise.
\subsection{Speech Intelligibility in Bits}
\label{subsec:SIIB}
\begin{figure}[!t]
    \centering
    \includegraphics[width=3.5in]{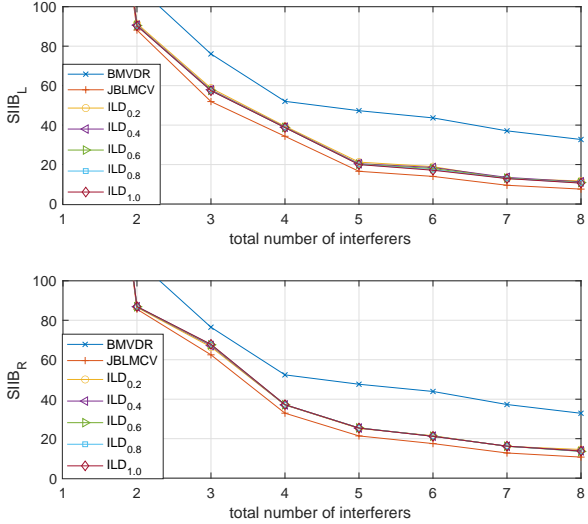}
    \caption{Speech intelligibility in bits index for the BMVDR, the JBLCMV and five different enhancements of the proposed method.}
    \label{fig:siib}
\end{figure}

The SIIB index uses the mutual information between the clean signal and the degraded signal to assess intelligibility. The algorithm incorporates the time-frequency dependencies in the speech signal and was found to be effective for speech degraded by noise and processed by enhancement algorithms \cite{van2017instrumental}. A higher bit index represents higher intelligibility whereas a lower bit index represents lower intelligibility. In this section, the BMVDR and the JBLCMV are compared with the different enhancements of the proposed method using the SIIB measure. The results can be seen in Fig. \ref{fig:siib} \textcolor{black}{for the signals at the left and the right ear that are represented as $SIIB_L$ and $SIIB_R$ respectively}. The BMVDR has the highest SIIB metric with 13 bits difference compared to the proposed methods and the JBLCMV has the lowest SIIB metric with 5 bits difference compared to the proposed methods at four interferers. 

The proposed methods share a \textcolor{black}{slightly better performance compared to} JBLCMV where the individual differences \textcolor{black}{between different enhancement amounts} can be neglected. It can be deduced that the applied low-frequency ILD enhancement does not affect intelligibility.  Since the low-frequency ITD cues are not preserved, the optimization problem has more degrees of freedom to do noise reduction. Hence, 5 bits of improvement has been observed compared to JBLCMV. It should be noted that the applied convex relaxation overestimates the optimal value, which is shown in (\ref{eq:overestimate}). Although in our analysis we observed that most of the time $W=ww^H$ holds, the noise reduction capability is still an overestimation of the original problem given in (\ref{eq:mainProblem}). In conclusion, the intelligibility of the signals is not disrupted with the low-frequency ILD enhancement. 
\subsection{Localization Performance}
\label{subsec:localizzzzzza}
ITD and ILD cues are the binaural cues that are important for horizontal localization. The ILD is defined as (\ref{eq:ILD}) and ITD is defined as (\ref{eq:ITD}). If the ILD at the input is the same as the ILD at the output, it can be said that the perfect preservation of the ILD has been achieved. However, if there is a mismatch between the input and output ILD, the ILD is not perfectly preserved. Depending on the frequency and the magnitude of the error, the horizontal location of the source might be different. The ILD error is defined to be 
\begin{align}
\label{eq:ILD_err}
    ILD_x^{err} = ||ITF_x^{out}|^2-|ITF_x^{in}|^2|\,,\\
    ILD_{u_i}^{err} =||ITF_{u_i}^{out}|^2-c|ITF_{u_i}^{in}|^2|\,.
\end{align}
where $c$ is the scaling factor that is chosen using (\ref{eq:DVFILD}) for each interferer $i=1\ldots r\leq r_{\rm{max}}$.

The interaural phase difference error is defined in (\ref{eq:IPD}). The $IPD^{err}$ is represented for the target and the interferer as
\begin{align}
\label{eq:IPD_err}
    IPD_x^{err}=\frac{\angle ITF_{x}^{out}-\angle ITF_{x}^{in}}{\pi }\,,\\
    IPD_{u_i}^{err}=\frac{\angle ITF_{u_i}^{out}-\angle ITF_{u_i}^{in}}{\pi }\,,
\end{align}
where $IPD^{err}\in [0,1]$. 

\begin{figure*}[!t]
    \centering
    \includegraphics[width=7.16in]{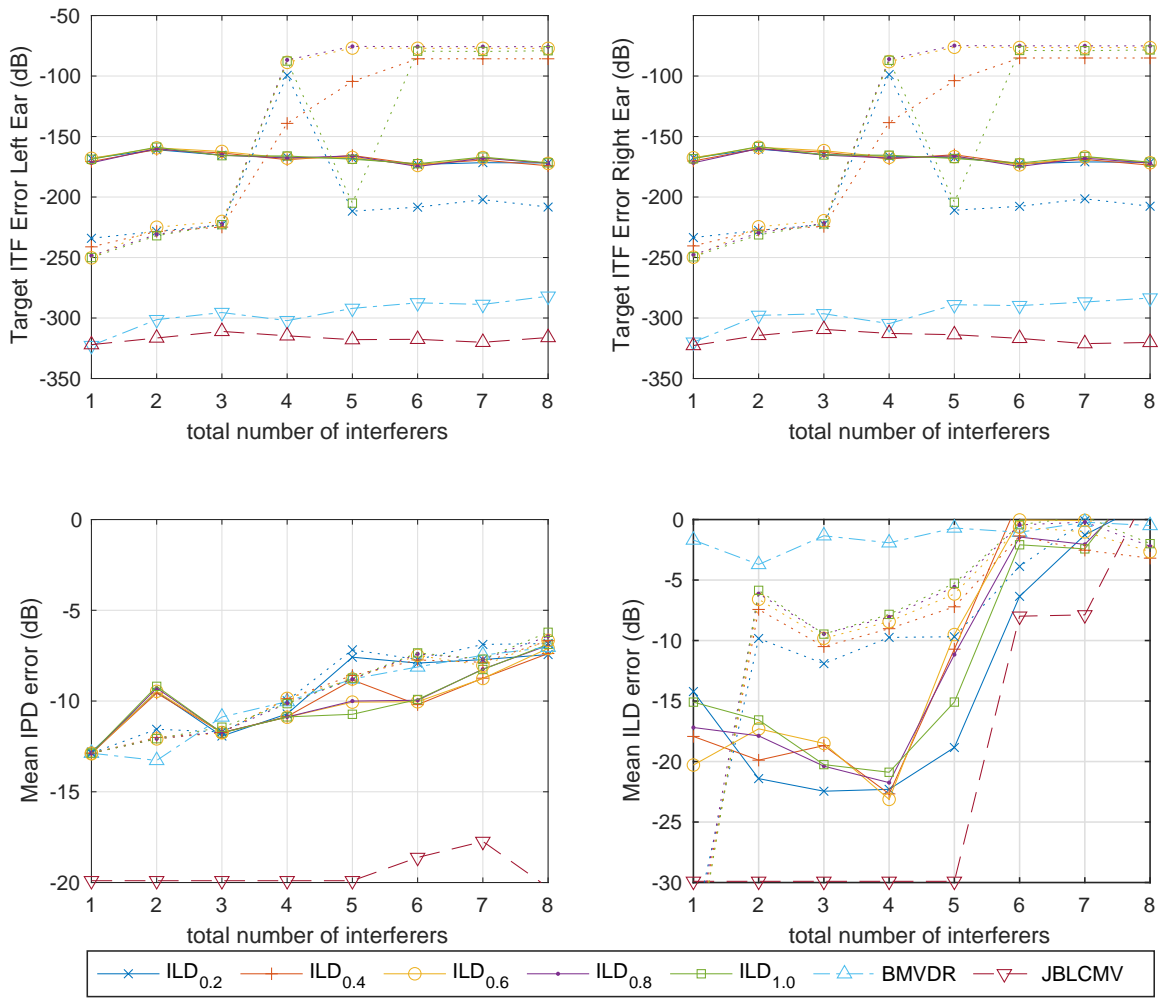}
    \caption{The interaural transfer function of the target and the interaural level and phase differences of the interferers plotted for the JBLCMV, the BMVDR and the five different enhancements of the two variants of the proposed method. The solid lines represent the optimization problem in (\ref{eq:OPTstep4}), the dotted lines represent the variant in (\ref{eq:OPTstep3}) for 5 different enhancements $ILD_{0.2}$, $ILD_{0.4}$, $ILD_{0.6}$. $ILD_{0.8}$ and $ILD_{1.0}$. The dashed line with triangle marker represent the JBLCMV method and the dot-dashed line with upside down marker represent the BMVDR method.}
    \label{fig:PerfTwoVariantLocalization}
\end{figure*}

The localization performance of the two variants of the original problem in (\ref{eq:mainProblem}), namely (\ref{eq:OPTstep3}) and (\ref{eq:OPTstep4}), will be examined for a different number of interferers. This investigation will be focused on $f<800$ Hz to analyze the performance of the proposed ILD enhancement beamformer performance separate from the JBLCMV. The reader is invited to \cite{AndreasAnalysis} for an analysis of the LCMV framework. For each frequency bin less than $800$ Hz, the error is calculated using (\ref{eq:ILD_err}) and (\ref{eq:IPD_err}). The results are averaged and transformed to the dB scale. The Fig. \ref{fig:PerfTwoVariantLocalization} includes the ITF error of the target source, ITD, and IPD errors of the interferers for the JBLCMV, the BMVDR, and the two variants of the proposed method with five different enhancements, which is $ILD_{0.2}$, $ILD_{0.4}$, $ILD_{0.6}$. $ILD_{0.8}$ and $ILD_{1.0}$. The IPD and ILD error of the JBLCMV was approximately $-130$ dB, which is represented by the lowermost line due to visualization purposes. \textcolor{black}{It can be seen that for all cases, the target signal is kept undistorted. The mean IPD error of the proposed beamformers is similar to BMVDR as none of them have any constraint about the preservation of the IPD, unlike the JBLCMV which has an error of $-130$ dB. The ILD error is higher when the interferer number is greater than $1$ when Problem 2 is used compared to the algorithm in Problem 3 due to the additional constraints.}
%\IEEEpubidadjcol
\section{Listening Experiment}
\label{sec:experiment}
\begin{figure}
    \centering
    \includegraphics[width=3.5in]{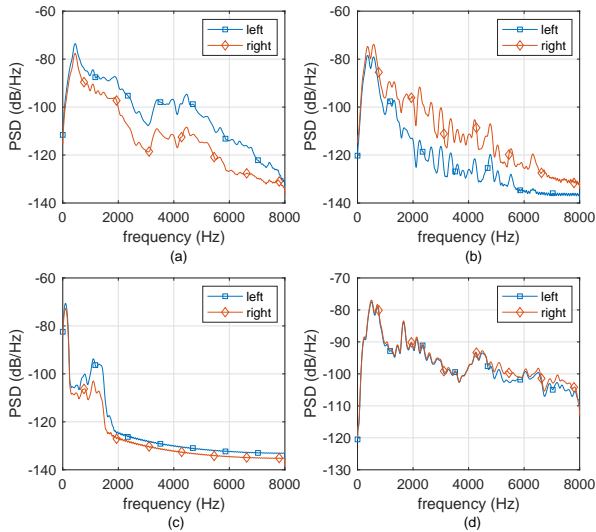}
    \caption{The power spectral density of the stimuli used in the experiment for the anechoic scene. The signals are (a) male speaker (b) piano tune (c) cellphone vibration and (d) female speaker.}
    \label{fig:psdofR2}
\end{figure}

An experiment is designed to assess the localization performance of the proposed method. 
The proposed method is compared to two reference methods, these are, the JBLCMV and the BMVDR. Each scenario consists of four signals.  Two of these signals are a male and a female speaker selected randomly from the TIMIT database \cite{garofolo1993darpa}. The third signal is a piano piece and the fourth signal is a cellphone vibration, which is low-pass filtered at 800 Hz. The female speaker is assigned as the target signal whereas, the three other signals are assigned as interferers. \textcolor{black}{The power spectral densities of the stimuli used in the experiment can be seen in Fig. \ref{fig:psdofR2} for the anechoic scene. The summary of the acoustic scenes can be seen in Table \ref{tab:ac}}.

\begin{table}[!t]
\renewcommand{\arraystretch}{1.3}
\caption{Summary of the acoustic scenes}
\begin{tabular}{|c|c|c|c|c|}
\hline
\multirow{2}{*}{\begin{tabular}[c]{@{}c@{}}Acoustic\\ scene\end{tabular}} & \multicolumn{4}{c|}{Source position}                                                                     \\ \cline{2-5} 
                                                                          & Female talker & Male talker & Piano tune & \begin{tabular}[c]{@{}c@{}}Cellphone\\ vibration\end{tabular} \\ \hline
\begin{tabular}[c]{@{}c@{}}Anechoic\\ (Scene 1)  \end{tabular}                                                               & 0             & -90         & 65         & -30                                                           \\ \hline
\begin{tabular}[c]{@{}c@{}}Office\\ (Scene 2)  \end{tabular}                                                                    & 0                            & 75          & 15        & -45                                                            \\ \hline
\end{tabular}
\label{tab:ac}
\end{table}

Each signal is processed with the BMVDR, the JBLCMV, and the three variations of the proposed method. The first variation preserves the natural low-frequency ILD cues up to 800 Hz while leaving the ITD cues unpreserved in this range. This method is abbreviated as $ILD_{1}$. The second variation artificially introduces ILDs in the low-frequency region up to 800 Hz with respect to $0.6$ m distance.  This method is abbreviated as $ILD_{0.6}$. The third variation artificially introduces ILDs in the low-frequency region up to 800 Hz with respect to $0.2$ m distance. This method is abbreviated as $ILD_{0.2}$. The enhancement scale is calculated according to (\ref{eq:DVFILD})
and applied per frequency bin. In all three variations, the JBLCMV beamformer is used for frequencies higher than 800 Hz.

The head-related transfer functions are selected using the middle and rear microphone recordings of the behind the ear (BTE) hearing aid \cite{kayser2009database}. There are two microphones on each of the left and right hearing aid totaling to four. Two different environments are used to understand the effect of reverberation. The first scene is the anechoic environment where the reverberation is minimal. The second scene is the office environment where the reverberation is greater than the anechoic environment. The anechoic head-related impulse response has a T60 of $50$ ms and the office head-related impulse response has a T60 of \textcolor{black}{$300$} ms where the early reflections are considered.

\begin{figure}
    \centering
    \includegraphics[width=3.5in]{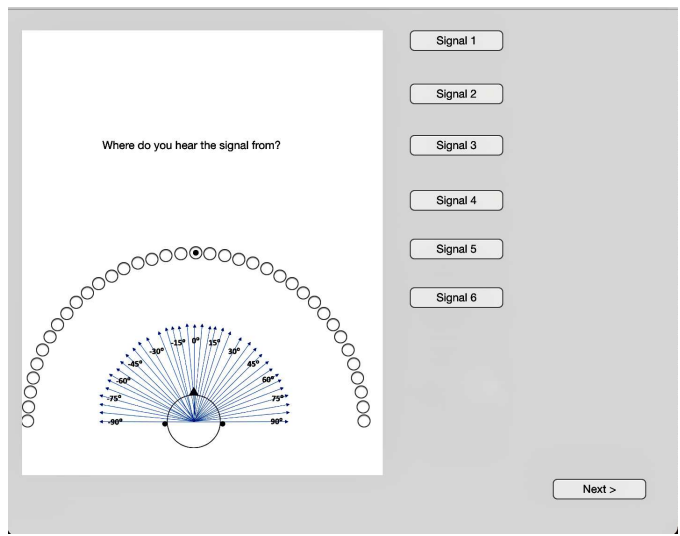}
    \caption{Graphical user interface for the listening experiment.}
    \label{fig:gui}
\end{figure}

An azimuth is assigned for each target and the interferer signal. The target signal is always kept at $0^\circ$ while the interferers are assigned to other angles in the frontal plane. \textcolor{black}{The graphical user interface depicted in Fig. \ref{fig:gui} is created. The user can click on a button as many times as they desire until they are confident with its location. There are $6$ buttons in one page, which correspond to JBLCMV, BMVDR, $ILD_{0.2}$, $ILD_{0.6}$, $ILD_{1.0}$ and the unprocessed version of the same signal. We consider a single source scenario where each button plays only one signal. The signals are 4 seconds long. The microphone self noise is simulated as an additive white Gaussian noise. Each recording has a $50$ dB SNR due to the self noise. The target and the interferers are processed such that they share the same power. The anechoic and the office scene recordings are played in a random order to prevent any bias. Each signal is presented twice. Since generic head related transfer functions are used, a mismatch is expected between the actual source locations and the interpreted locations. We included the unprocessed signals to account for this mismatch. The locations assigned to the unprocessed signals are used as a reference azimuth for the calculation of the localization error.} 

Participants are asked to attend the experiment online. This causes \textcolor{black}{a} few intricacies that need to be handled. Firstly, the participants are asked to do the test in a silent room to avoid the interference of other sounds. Secondly, the participants are informed to finish the whole experiment without changing the volume after getting used to it initially. This prevents any cues related to increased audibility. Last but not least, a guideline is prepared to make sure that the headphones are placed correctly and the system is working. 
\subsection{Results}
In total, 21 people aged between $20-27$ participated in the experiment. None of the participants had any reported hearing problems. Since it was not possible to control the environment or the tools that the participants were using, a variance test is done to find if there are outliers. The variance is calculated using the results of the unprocessed case and its repetitions. If the participants assigned significantly different angles to the same unprocessed signals (twice repeated), they are considered outliers. Only one participant is found to be three standard deviations higher than the others. This participant's result has been omitted from the analysis reducing the number of participants to 20.

\begin{figure}[!t]

    \centering
    \includegraphics[width=3.5in]{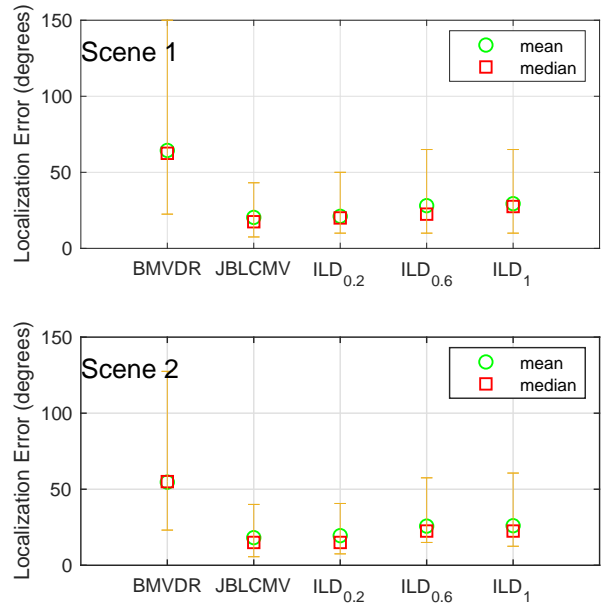}
    \caption{Mean, median, 0.25 and 0.75 quantiles of the localization error of all the participants at two different scenes, averaged over the interferers.}
    \label{fig:AllInterferer}
\end{figure}

The averaged results of the experiment can be seen in Fig. \ref{fig:AllInterferer}.
Two-way ANOVA with repeated measures is used with source type as within-group factors and different beamforming algorithms as between-group factors
to compare the localization performances of the beamformers with the spatial cue preservation constraint, that is, the JBLCMV and the proposed method with different enhancements. Each participant chooses the angle of a sound source twice independently. The answers are averaged to reduce the variance per participant. Three \textcolor{black}{hypotheses} have been \textcolor{black}{postulated}. The first hypothesis \textcolor{black}{denoted as} $H^a_0$ examines if there is a significant difference between the methods: the JBLCMV, $ILD_{0.2}$, $ILD_{0.6}$, and $ILD_{1.0}$. This hypothesis assesses how close the JBLCMV and the proposed methods are with respect to localization performance. Since the JBLCMV preserves both the ILD and ITD cues in the whole spectrum, it is expected to perform the best. If there is a significant difference between the JBLCMV and the proposed methods, it can be deduced that the localization is hindered due to the lack of ITD preservation in the low-frequency region. The second hypothesis, \textcolor{black}{denoted as} $H^b_0$, examines if any of the sound sources have a significantly different localization outcome. The third hypothesis, \textcolor{black}{denoted as} $H^c_0$, examines the interaction between the beamformers and the sound sources. The former hypothesis assesses if the localization performance is different for a low-pass filtered sound, speech sound or a broadband piano piece. The latter hypothesis, on the other hand, explores if any of the beamformer performances are correlated with the sound source type. The summary of the results is given in Table \ref{tab:anova}.

\begin{table}[!t]
\renewcommand{\arraystretch}{1.3}
\centering

\caption{Summary of the ANOVA results}
\begin{tabular}{|c|c|c|c|}
\hline
Hypothesis                               & Acoustis Scene & F     & Prob\textgreater{}F \\ \hline
\multirow{2}{*}{$H^a_0$} & Scene 1       & 4.6   & 0.0038              \\ \cline{2-4} 
                                         & Scene 2         & 3.01  & 0.0300              \\ \hline
\multirow{2}{*}{$H^b_0$} & Scene 1       & 3.46  & 0.0330              \\ \cline{2-4} 
                                         & Scene 2         & 12.26 & \textless{}0.0001   \\ \hline
\multirow{2}{*}{$H^c_0$} & Scene 1       & 0.46  & 0.8382              \\ \cline{2-4} 
                                         & Scene 2         & 0.94  & 0.4620              \\ \hline
\end{tabular}
\label{tab:anova}
\end{table}

For the anechoic scene, the first hypothesis $H^a_0$ is rejected due to [$F(3,228)=4.6; p\sim=0.0038$], which suggests that at least one of the algorithms is significantly different from the others. A follow-up test is done at the end of the section to understand which of the methods are significantly different.

For the office environment, the results are similar to the first scene. $H^a_0$ is rejected due to [$F(3,228)=3.01; p
=0.03$]. This suggests that the beamformer algorithms the JBLCMV, $ILD_{0.2}$, $ILD_{0.6}$, and $ILD_{1.0}$ perform differently also for scenarios with higher reverberation. The second hypothesis is rejected as [$F(2,228)=12.26; p<0.0001$]. This suggests that at least one of the sources is significantly different from the other sources. The last hypothesis $H^c_0$ is accepted as [$F(6,228)=0.94; p=0.46$]. There is no significant interaction between any of the algorithms and the type of the source.

In both scenes, it has been found that there is a significant difference between the performance of the JBLCMV and the proposed method with different enhancements. To understand which ones are significantly different from the JBLCMV, post-hoc analysis is done using Tukey's test \cite{tukeys}. The summary of the results can be seen in Table \ref{tab:tukey}.

\begin{table}[!t]
\caption{Post-hoc analysis of the sources using Tukey's test}

\renewcommand{\arraystretch}{1.3}
\centering
\begin{tabular}{|c|c|c|c|}
\hline
\multicolumn{2}{|c|}{\begin{tabular}[c]{@{}c@{}}Beamforming\\ Algorithms\end{tabular}} & \begin{tabular}[c]{@{}c@{}}Acoustic\\ Scene\end{tabular} & p-value              \\ \hline
\multirow{2}{*}{$ILD_{0.2}$}               & \multirow{2}{*}{$ILD_{0.6}$}              & Scene 1                                                 & 0.0822               \\ \cline{3-4} 
                                           &                                           & Scene 2                                                   & 0.2038               \\ \hline
\multirow{2}{*}{$ILD_{0.2}$}               & \multirow{2}{*}{$ILD_{1.0}$}              & Scene 1                                                 & 0.0242               \\ \cline{3-4} 
                                           &                                           & Scene 2                                                   & 0.1640               \\ \hline
\multirow{2}{*}{$ILD_{0.2}$}               & \multirow{2}{*}{JBLCMV}                   & Scene 1                                                 & \textgreater{}0.9999 \\ \cline{3-4} 
                                           &                                           & Scene 2                                                   & 0.9964               \\ \hline
\multirow{2}{*}{$ILD_{0.6}$}               & \multirow{2}{*}{$ILD_{1.0}$}              & Scene 1                                                 & 0.9688               \\ \cline{3-4} 
                                           &                                           & Scene 2                                                   & 0.9995               \\ \hline
\multirow{2}{*}{$ILD_{0.6}$}               & \multirow{2}{*}{JBLCMV}                   & Scene 1                                                 & 0.0793               \\ \cline{3-4} 
                                           &                                           & Scene 2                                                   & 0.1302               \\ \hline
\multirow{2}{*}{$ILD_{1.0}$}               & \multirow{2}{*}{JBLCMV}                   & Scene 1                                                 & 0.0232               \\ \cline{3-4} 
                                           &                                           & Scene 2                                                   & 0.1020               \\ \hline
\end{tabular}
\label{tab:tukey}
\end{table}

The Tukey's significance test for scene 1 shows that there is a significant difference between the JBLCMV and $ILD_{1.0}$ with $p=0.0232$. Moreover $ILD_{0.2}$ and $ILD_{1.0}$ are also found to be significantly different with $p=0.0242$.  There is no significant difference captured among the other methods. As both the JBLCMV and $ILD_{0.2}$ are significantly different from $ILD_{1.0}$, it can be deduced that the loss of information induced by the low-frequency ITD cues has been overcome by the enhancement of the ILDs in the low-frequency range.

Tukey's significance test for scene 2 however, does not show any significant differences among the proposed methods although a similar trend can be observed. The p-value between the JBLCMV and $ILD_{1.0}$ is $0.1020$, between $ILD_{0.2}$ and $ILD_{0.6}$ is $0.1640$ and between $ILD_{0.2}$ and the JBLCMV is $p=0.9964$. The significant improvement of horizontal localization performance that is observed for the anechoic scene is not observed for the reverberant scene.

Since no interaction has been found between the proposed beamformers and the source type, further analysis has not been conducted for these cases.

\section{Discussion}
This project \textcolor{black}{aims} to understand if the low-frequency ITD information can be exchanged with the ILD information in the same range. The intelligibility and the localization performance of the proposed method have been theoretically examined in Section \ref{sec:performance}. A psychoacoustic analysis is done using a listening experiment to understand if indeed there is any improvement in localization performance as reported in the literature.

The proposed method targets hearing-impaired people with a low temporal fine structure processing ability. There are several methods in the literature such as \cite{mathew2020measuring}, which focuses on measuring the TFS sensitivity. By understanding the auditory capabilities of the hearing impaired, those who will benefit from the proposed method can be selected. In addition to this, cochlear hearing loss is known to cause a deficiency in temporal fine structure processing \cite{hopkins2008effects}. The hearing-impaired people with cochlear hearing loss has been known to be utilizing the envelope time differences and the level differences in the high frequencies. This makes the proposed method, which preserves the phase and the levels in the high frequency while replacing the low-frequency ITD information with ILD a suitable method for those who suffer from cochlear hearing loss. 

\textcolor{black}{In our experiment, we used normal hearing subjects to understand if the low-frequency ILD cues can be used to overcome the lack of ITD cues in the same region. In an attempt to imitate the deficiency of TFS processing, we have introduced low-frequency ITD errors during beamforming. From the binaural study given in Fig. \ref{fig:binauralbeat}, we have found out that some of the hearing-impaired people are able to hear some of the low-frequency ITD cues. We have created a scenario where the frequencies below $800$ Hz do not contain reliable ITD information, whereas the frequencies above $800$ Hz do. With the introduction of near-field ILD cues below $800$ Hz, we have observed improved localization performance. The hearing-impaired people share a similar condition where six of the attendees were able to hear some of the ITD cues in the low-frequency region whereas, nine of them could not hear any at all. We believe the listening test on normal hearing people generalizes to the hearing-impaired people, where the attendees were asked to localize a sound source with deficient ITD cues. However, the same performance that is observed for normal hearing people might not be observed for hearing-impaired people. This should be further tested.}

The results in Section \ref{subsec:SIIB} suggest that the low-frequency ITD information is not directly related to intelligibility or its effect below $800$ Hz can be compensated by the information that exists above $800$ Hz. The intelligibility metric SIIB shows that the enhancement of the low-frequency ILDs does not disrupt the intelligibility of the signals as the proposed method has better intelligibility compared to JBLCMV. In the literature, a high spatial release from masking is reported when the low-frequency ILDs are introduced. This effect \textcolor{black}{has} not been observed in the intelligibility metrics and should be examined in future work.

Section \ref{subsec:localizzzzzza} covers the localization performance of the proposed method compared to the BMVDR and the JBLCMV using the binaural cues. It can be seen that Problem 3 given in (\ref{eq:OPTstep4}) has overall better performance. The desired low-frequency ILDs can be reached with less error compared to Problem 2 given in (\ref{eq:OPTstep3}). In addition to the interferer localization properties, the target is left distortionless for any number of interferers as expected. There is no objective localization measure that assesses the localization ability of the hearing impaired that includes auditory deficiencies. For this reason, the binaural cues preservation performance is examined in an attempt to show the performance of the optimization problem at solving the original problem. 

The results in Section \ref{sec:experiment} suggest that it is possible to replace the ITD information using low-frequency ILDs. In the averaged results, which can be seen in Fig. \ref{fig:AllInterferer}, the localization performance of $ILD_{0.2}$ is close to the JBLCMV. On the other hand, $ILD_{0.6}$ and $ILD_{1.0}$ have a higher variance and a higher mean error compared to $ILD_{0.2}$ and the JBLCMV.

\textcolor{black}{As the magnitude of the ILD cues in the low-frequency region increases, the localization performance improves. This behaviour can be explained by the decrease in the audible angle with increasing ILD magnitude. It might be possible to differentiate the difference between the $90^\circ$ and $60^\circ$ when enhancement with respect to $0.2$ m is applied whereas, it might not be possible when enhancement with respect to $0.6$ m is applied. In addition, the enhancement amount is expected to be dependent on the sensitivity of the hearing-impaired listener to ILD differences. On the other hand, a higher enhancement might impair the distance perception of the hearing-impaired user since the low-frequency ILD cues are mainly used as distance cues. This relation will be examined in the feature work to understand the trade-off.}

\section{Conclusion}
In this research project, a step towards manipulating the acoustic scene according to the hearing loss has been taken. The low-frequency ITD cues, which some of hearing-impaired people have shown problems of processing, are transformed into ILD cues using a near-field transformation. The listening test suggests that the localization ability of the normal listeners can be improved when enhancement with respect to $0.2$ m has been applied for the anechoic scene. This amount is expected to be user-dependent and can be tailored using psychoacoustic analysis. An intelligibility and localization experiment on hearing-impaired people with deficient TFS processing is left for future work. 

\ifCLASSOPTIONcaptionsoff
  \newpage
\fi

\end{document}